\PassOptionsToPackage{dvipsnames}{xcolor}
\documentclass[sigconf]{acmart}
\AtBeginDocument{%
  }

\copyrightyear{2026}
\acmYear{2026}
\setcopyright{cc}
\setcctype{by-nc-nd}
\acmConference[CHI '26]{Proceedings of the 2026 CHI Conference on Human Factors in Computing Systems}{April 13--17, 2026}{Barcelona, Spain}
\acmBooktitle{Proceedings of the 2026 CHI Conference on Human Factors in Computing Systems (CHI '26), April 13--17, 2026, Barcelona, Spain}
\acmPrice{}
\acmDOI{10.1145/3772318.3791180}
\acmISBN{979-8-4007-2278-3/2026/04}

\usepackage[shortlabels]{enumitem}
\usepackage{xspace}

\usepackage[capitalize]{cleveref}

\newcommand{\GMNEXL}{P1.1\xspace}
\newcommand{\GYGLNT}{P1.2\xspace} 
\newcommand{\MJCEWA}{P1.3\xspace} 
\newcommand{\OQUJBJ}{P1.4\xspace} 
\newcommand{\RZWGHO}{P1.5\xspace} 
\newcommand{\WGAJTB}{P1.6\xspace} 
\newcommand{\ZIKDKJ}{P1.7\xspace} 
\newcommand{\BLQUQS}{P2.1\xspace} 
\newcommand{\DDMHLY}{P2.2\xspace} 
\newcommand{\EZFYQG}{P2.3\xspace} 
\newcommand{\GUKOJD}{P2.4\xspace}
\newcommand{\JPBRXV}{P2.5\xspace} 
\newcommand{\KOLLDL}{P2.6\xspace} 
\newcommand{\RTQBCI}{P2.7\xspace} 
\newcommand{\WYCXDJ}{P2.8\xspace} 
\newcommand{\DPYKRI}{P3.1\xspace} 
\newcommand{\LVAZBN}{P3.2\xspace} 
\newcommand{\SFWPFB}{P3.3\xspace} 
\newcommand{\SRXTXD}{P3.4\xspace} 
\newcommand{\TMZZKH}{P3.5\xspace} 
\newcommand{\UWADUD}{P3.6\xspace} 
\newcommand{\YSDDVC}{P3.7\xspace}

\begin{document}
\title{%
Relational Dissonance in Human-AI Interactions: The Case of Knowledge Work
}

\author{Emrecan Gulay}
\orcid{0000-0002-7064-8115}
\affiliation{%
  \institution{Aalto University}
  \city{Greater Helsinki}
  \country{Finland}}

\author{Eleonora Picco}
\orcid{0009-0004-9105-2064} 
\affiliation{%
  \institution{Aalto University}
  \city{Greater Helsinki}
  \country{Finland}}

\author{Enrico Glerean}
\orcid{0000-0003-0624-675X}
\affiliation{%
  \institution{Aalto University}
  \city{Greater Helsinki}
  \country{Finland}}

\author{Corinna Coupette}
\orcid{0000-0001-9151-2092}
\affiliation{%
  \institution{Aalto University}
  \city{Greater Helsinki}
  \country{Finland}}

\renewcommand{\shortauthors}{Gulay et al.}

\begin{abstract}
When AI systems allow human-like communication, they elicit increasingly complex relational responses. Knowledge workers face a particular challenge: They approach these systems as tools while interacting with them in ways that resemble human social interaction. To understand the relational contexts that arise when humans engage with anthropomorphic conversational agents, we need to expand existing human-computer interaction frameworks. Through three workshops with qualitative researchers, we found that the fundamental ontological and relational ambiguities inherent in anthropomorphic conversational agents make it difficult for individuals to maintain consistent relational stances toward them. Our findings indicate that people's articulated positioning toward such agents often differs from the relational dynamics that occur during interactions. We propose the concept of \emph{relational dissonance} to help researchers, designers, and policymakers recognize the resulting tensions in the development, deployment, and governance of anthropomorphic conversational agents and address the need for \emph{relational transparency}. 

\end{abstract}


\begin{teaserfigure}
  \centering
  \includegraphics[width=0.800
  \textwidth]{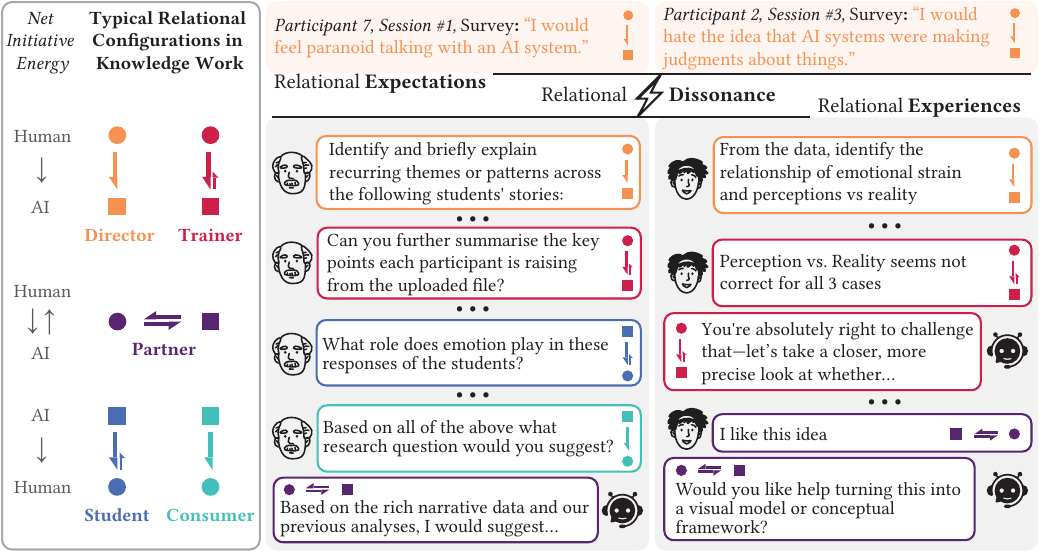}
  \caption{\textbf{Core components of the relational-dissonance framework, illustrated on data collected from knowledge workers.}
  }\label{fig:teaser}
  \Description{%
  Three-column figure illustrating the relational-dissonance framework. In the left column, we show five typical relational configurations in knowledge work, along with their net initiative energy: Director (human gives unidirectional commands to AI), Trainer (human guides with bidirectional feedback), Partner (equal bidirectional exchange), Student (AI guides human), and Consumer (human passively receives from AI). In the middle and right column, we illustrate the relational-dissonance concept, showing misalignment between relational expectations and relational experiences based on survey responses and chat excerpts from two study participants (P1.7 middle, P3.2 right) showing dialogue exchanges with AI systems. For both participants, the survey responses indicate Director-type expectations, but the actual chat interactions demonstrate multiple shifts toward configurations with less human control.
  } 
\end{teaserfigure}


\maketitle
\clearpage

\section{Introduction}\label{sec:introduction}

\emph{``It adapts to what you want it to be, but it is not really any one thing.''}

This insight from a participant in our study (\EZFYQG) captures a fundamental aspect of human engagement with \emph{anthropomorphic conversational agents} \cite{Peter2025}, i.e., Artificial Intelligence (AI) systems that mimic human communication (hereafter: \emph{AI~systems}). 
Humanity has long been fascinated by creating machines mirroring its own image \cite{Cave2018},
and today's AI systems have achieved sophisticated communicative capabilities \cite{Bonan2023}.
However, as humans increasingly rely on AI systems to complete complex analytical tasks \cite{fernandez2020, Gao2023},
the nature of what they are engaging with remains elusive,
and the interactions that unfold challenge our established conceptual categories. 
In particular, 
while existing work in human-computer interaction (HCI) helps us understand human-computer social engagement \cite{Nass1994, Nass2000, Gambino2020, Pentina2023},
we still lack the analytical language to describe what happens when people engage with anthropomorphic conversational agents. 
The need for this language becomes especially apparent in the context of knowledge work, 
where AI systems are primarily discussed as practical tools for problem solving or decision support \cite{Mossbridge2024, McGee2025, tominaga2025}---%
in line with occupational expectations of human agency and control, 
yet contradicting the opaque and unpredictable nature of generative AI \cite{Hassija2024, Zhao2024}. 
More fundamentally,
AI systems operate in ways that challenge our fundamental understanding of what we are engaging with,
regardless of how well we understand their mechanisms.
As summarized by \Cref{fig:teaser}, in this work, we propose the concept of \emph{relational dissonance} as a step toward filling this gap. 

\textbf{Defining relational dissonance.}
\emph{Relational dissonance} describes the divergence between (1) an individual's ontological and relational framing of AI systems and (2) the relational dynamics that emerge as their interaction with these systems unfolds.

In our conception,
relational dissonance captures the difficulty of keeping stable answers to two questions about AI:
``What is AI?'' (\emph{ontological ambiguity}) and ``What is AI \emph{to me}?'' (\emph{relational ambiguity}). 
The resulting dynamics can lead to a lack of alignment between individuals’ articulated accounts (\emph{explicit dimension}) and enacted relationships (\emph{implicit dimension}),
which remains mostly unnoticed during an individual's interaction with a system. We understand relational dynamics as observable qualities in human-AI interactions,
including the roles enacted and the social tone of the exchange. 
Building on recent HCI research \cite{Kegel2025, Fu2025},
we use the term \emph{relational} to differentiate socially complex interactions from purely transactional exchanges. 
Our focus lies on the live, momentary dynamics of a person interacting with what Turkle et al. \cite{Turkle2006RelationalCybercompanionship} call a \emph{relational artifact}, i.e., an entity that invites partnership-like responses without having subjectivity or needs. 
To the best of our knowledge,
our definition of relational dissonance creates a novel theoretical construct specific to human-AI interactions,
completely distinct from a prior narrow appearance of the term in adolescent psychology \cite{Bond2014}.

\textbf{Situating relational dissonance.}\quad 
The term \emph{dissonance} originates from music theory \cite{Levin1972},
where it describes sounding apart relative to an expected line. 
In line with its semantics, we use this term to capture a built-in tension in human-AI interactions.
In our framework,
we employ `consonance' and `dissonance' as descriptive tools.
We do not position consonance as an ideal state to achieve,
nor treat dissonance as a problem requiring resolution.
Instead,
we understand relational consonance as the alignment between the explicit and the implicit dimension of an individual's interaction with AI systems. 
Relational dissonance, then, refers to the lack of alignment between these dimensions.
When we identify dissonance,
we do not imply subjective conflict,
cognitive error,
or the existence of a `correct' way to relate to AI systems.
Rather,
we use relational dissonance as a lens to illuminate the fundamental ambiguity inherent in human-AI interactions and the relational shifts that occur as these interactions unfold. 

Similar to Festinger's \emph{cognitive dissonance} \cite{Festinger1957},
relational dissonance operates partially outside our conscious awareness.
Cognitive dissonance has been defined as a felt intrapsychic conflict between incompatible cognitions that people attempt to resolve. 
The key difference from this concept is that the locus of tension in relational dissonance is not within an individual's psyche;
rather,
it is in the structure of the interactive human-AI interface.
Individual strategies alone,
such as changing attitudes or behaviors,
cannot resolve AI's inherent ambiguity. 
The instability keeps regenerating due to AI's unstable ontological and relational status,
regardless of what people think or feel.
However,
relational dissonance is not neutral;
in contexts such as knowledge work,
it carries real consequences that range from the psychological (e.g., fearing loss of agency) and relational (e.g., overdependence concerns) to the professional (e.g., bearing the accountability for the work output). 
People continuously navigate this built-in tension without recognizing the shifts in their relational roles and postures.

The \emph{intention–behavior gap}, a well-documented phenomenon denoting the disconnection between people’s stated intentions and what they do \cite{Sheeran2002},
can coexist with relational dissonance, 
but, like cognitive dissonance, its locus remains internal to the individual. 
In relational dissonance,
the lack of alignment occurs not between individual intentions and actions,
but rather between a person's \emph{explicit} ontological and relational framing of AI systems and the \emph{implicit} relational dynamics that unfold during the course of their interactions with these systems.

\textbf{Observing relational dissonance.}\quad
We developed relational dissonance as an analytical construct through three workshops with a total of 22 qualitative researchers, 
conducted to answer the following research question:

\textbf{RQ}\quad \emph{What forms of relational dynamics emerge when qualitative researchers engage with AI systems?}

We focused specifically on qualitative research because it exemplifies knowledge work that comes with high social expectations of human agency and control: 
Qualitative researchers engage in deep interpretive work that involves constructing meaning and critically reflecting on interpretive choices \cite{Flick2023, Braun2022}, and any external influence on this process can be viewed as a threat to research validity.
When AI systems enter the qualitative research process,
they introduce a new relational other \cite{dehnert2023ai},
and meanings become co-constructed through repeated cycles of prompting,
interpretation,
and refinement.
The qualitative research process thus offers a window into how relational dynamics arise when AI systems are integrated into complex cognitive work that has traditionally been exclusive to human researchers.

\textbf{Key contributions and paper organization.}
In this work, we make three key contributions.
\begin{enumerate}[label=\bfseries C\arabic*,nosep]
    \item We introduce relational dissonance as a novel analytical construct to characterize human-AI relational dynamics. 
    \item We offer a multilayer workshop design for studying human-AI relational dynamics.
    \item We provide nuanced insights into human-AI relational dynamics in the context of qualitative research. 
\end{enumerate}
To develop these contributions,  
we begin by connecting our study to related work on human-AI interaction (\cref{sec:related}).
Next, we describe our methodological approach,
detailing our workshop design and our analytical process (\cref{sec:methods}).
We then present our findings,
examining both the individual and the collective dimensions of relational dissonance that emerged in our study (\cref{sec:findings}),
before discussing how our findings enhance our understanding of human-AI relational dynamics in the context of knowledge work and highlighting the need for \emph{relational transparency} (\cref{sec:discussion}).
Finally,
we acknowledge the limitations of our study (\cref{sec:limitations}) and outline directions for future work (\cref{sec:conclusion}).

\section{Related Work: Understanding AI Ontology and Relationality in Human-AI Interactions}\label{sec:related}

Researchers have tried to comprehend AI systems from various angles.
At the \emph{ontological level},
research concentrates on understanding with what,
or with whom,
people are interacting \cite{Etzrodt2021}.
This body of work largely emphasizes the instrumental use of AI, 
foregrounding efficiency and other benefits,
often with limited attention to the individual.
Within the managerial literature,
most studies examine the impact of AI technologies on a range of workplace outcomes \cite{budhwar2023, PEREIRA2023},
highlighting their positive effects on individual performance, satisfaction,
commitment,
and employee experiences \cite{Ashish2023},
which,
in turn,
shape firm-level outcomes \cite{LEE2023, Malik2022}. 

Similarly, 
\emph{HCI~experts} have focused on addressing 
people’s acceptability and acceptance of text-based chatbots \cite{Koluluori2022}, mapping users’ judgments and attitudes \cite{wang2025, Yuan2025}, as well as on understanding people’s experience with text-based chatbots \cite{Ma2024}, capturing users’ expectations \cite{Malkomsen2024}, perceptions \cite{Fan2025}, conversational issues \cite{Jeong2025}, and a variety of effects, such as satisfaction, engagement, trust, and emotional impacts \cite{lee2025, Lutzelschwab2025, Genc2025}. 
Within organizational and information-systems research,
the \emph{sociotechnical~perspective} views technology both as the product of strategic choice and social action and as an external force with effects on organizational structures and humans \cite{orlikowski1992duality}.
Scholars in \emph{work and organizational psychology} (WOP) have argued that the majority of studies on human-AI collaboration draw a sharp distinction between humans and AI,
seeing AI either as a tool supporting tasks and work or as a medium for collaboration between groups in organizations \cite{anthony2023collaborating, bankins2024multilevel, le2024emerging}. 
In parallel,
some management research adopts a more critical perspective on the consequences of introducing AI into organizational decision-making and collaboration processes.
In addition to underscoring the potentially adverse implications of AI use for employees (e.g., increased monitoring and control) and its broader effects on those affected by AI-supported decisions (such as unwarranted biases),
these studies have shown how AI technologies mediate organizational practices like knowledge sharing and knowledge hiding \cite{Perez2021, liu2024},
and how the tools,
applications,
and outcomes of these technologies reinforce gendered role stereotypes \cite{Craiut2022, Nadeem_Marjanovic_Abedin_2022, Shrestha2022}. Within the WOP literature, alternative proposals position AI as an active counterpart within a system of interactions---%
to better understand its role in organizational processes and to anticipate the unintended consequences that may arise when all stakeholders work ‘with’ AI \cite{anthony2023collaborating}. Such a systems perspective also lends itself to examining the nature of the relationships among different actors, including people and AI.

At the \emph{relational level}, a growing body of HCI literature focuses on humanness in human-chatbot interaction, pointing out both how human aspects are emulated \emph{by} the system and how people perceive humanness \emph{in} the system \cite{rapp2021human}.
Not without difficulties in defining it,
this literature understands humanness as referring to the qualities, experiences,
and behaviors that are (uniquely or predominantly) associated with being human (e.g., communication, social connection, intolerance, empathy, among many others) \cite{hond2025}.
Some studies \cite{Etzrodt2021} argue that subjectivity should be located in personhood rather than humanness,
since `human' denotes biological belonging to a species,
whereas being a person involves the acknowledgment of personhood and is therefore culturally determined \cite{Hubbard2011DoAndroidsDream}. 
Along this line,
the \emph{human–AI communication} (HAIC) literature frames AI systems as communicative others not only because they interact with human users through language-based or sensory modalities \cite{dehnert2023ai, shaikh2023artificially},
but also because they are linguistically conceptualized in this way.
Here,
AI is constituted as `other' through its relation to the human,
not only by inherent ontological characteristics or properties \cite{dehnert2023ai}.
Similarly,
it has been argued that we must move beyond the notion of anthropomorphism as the attribution of human qualities to nonhuman systems by individuals \cite{Waytz2010}---%
toward a perspective that embeds anthropomorphic qualities in the machine itself within the relationship~\cite{Peter2025}. 
Besides anthropomorphism,
scholars have used other theories to explain how,
when engaging with computers, people tend to apply the same social rules they use with other humans. The \emph{Computers as Social Actors} (CASA) paradigm showed that people exhibit social behaviors when interacting with machines, even when they do not consciously believe that the computers are human-like \cite{Nass1994}. More recently, CASA has been expanded to suggest that humans may apply distinct human–media social scripts in these interactions, rather than human-human social scripts  \cite{Gambino2020}. 
 Within the psychology literature,
 human-AI relationships have also been proposed to satisfy fundamental human psychological needs (i.e., comfort and emotional security, self-identity formation, and personal efficacy) \cite{wan2021anthropomorphism}.
 For example,
 Winnicott’s concept of `transitional object,'
 an entity occupying the ambiguous space between self and other \cite{Winnicott2016},
 offers a powerful lens for conceptualizing AI systems. 
 Several \emph{interpersonal relationship theories} have been employed to explain human–AI relationship-development mechanisms \cite{Pentina2023}, such as social penetration theory \cite{altman1973social} and attachment theories \cite{west1994patterns, Pentina2023}. Within the domain of media studies, the idea of parasocial interaction describes one-directional, mediated relations with favored media characters, which has been extended to AI interactions \cite{auter2000development}. In the field of computer-mediated communication, the concept of social presence has been used to explain the feeling of `another being, either living or synthetic', that appears to react to you  
\cite{heater1992being, Pentina2023}. Several other theories also emphasize mediated interactions carried out specifically through written language (e.g., \cite{joinson2001self, tidwell2002computer, valkenburg2009effects, walther1996computer}). Yet, these theories alone do not fully capture the complex nature of human-AI interaction~\cite{rapp2021human}. Rather, the fragmentation of approaches complicates understanding the connection between AI ontology and relationality in human-AI interaction.

Prior work has revealed that anthropomorphic conversational agents are assimilated as personified things \cite{Etzrodt2021}.
Accordingly, the starting point of our study is the observation that regardless of AI's ontological status, relationality arises in human-AI conversations, constructed through the use of language \cite{coeckelbergh2010robot, dehnert2023ai}. In this sense, human-AI relationships are neither fully human-directed nor AI-determined---rather, they are co-created through human-AI interaction \cite{Raikov2022}. 
 However, the multifaceted and complex nature that characterizes interactions between human beings may be intensified or triggered by the interaction with AI systems, leading to ambiguities and a variety of effects \cite{hond2025}.
 For example, in the psychology
 literature, \emph{Uncanny Valley
 Theory} argues that perceptual
 ambiguity in identifying human
 like objects tends to evoke
 adverse emotional responses
 \cite{mori2012uncanny}.
At the same time, persistent instability in the personality measurement of Large Language Models (LLMs) indicates that current LLMs lack the architectural basis for genuine behavioral consistency \cite{tosato2025}.
With LLMs, users and relationships are changing rapidly \cite{Gambino2020, zimmerman2024human}, and it is crucial to investigate how the pervasive human-likeness of AI affects people’s interaction experiences, while also addressing its benefits, risks, and ambiguities. This involves asking what a relationship with AI systems looks and feels like \cite{dehnert2023ai}, and how AI’s ontology is connected to its relational effects. 
Addressing these questions in the context of knowledge work carries far-reaching implications, given the central role of relationships in shaping work practices. Human-AI interactions are redefining the meaning of activities and dignified work \cite{verboom2025perceptions}, reshaping skills, and having consequences that extend beyond the workplace to influence the broader trajectory of societal history and human evolution \cite{brinkmann2023machine}. Furthermore, in the context of knowledge work, anthropomorphizing nonhuman agents can particularly serve the basic human motivation to make sense of an otherwise uncertain environment \cite{Waytz2010}.

\section{Methods}\label{sec:methods}

To examine the complex relational dynamics between qualitative researchers and AI systems,
we adopted a multilayered workshop protocol capturing both the explicit accounts researchers provide of AI systems and the implicit relational dynamics unfolding in their interactions with these systems (\cref{subsec:study-design}).
As further detailed in \cref{subsec:analytical-framework}, we employed two complementary analytical frameworks to identify both individual engagement patterns (via interpretative phenomenological analysis, IPA) and collective sense-making processes (via thematic analysis, TA). 
IPA comprises a phenomenological component, which maps out participants’ lived experiences, and an interpretative component, which situates these experiential claims within the person’s broader context \cite{Larkin2006}. 
TA, in turn, focuses on identifying patterns of meaning (themes) within qualitative data, with the researcher making active interpretative choices in generating codes and constructing themes. 
Although TA is not tied to any particular theoretical framework, it aligns well with the assumptions of social phenomenology \cite{BraunClarke2012}. 
Conversely, IPA is flexible in its interpretative reach, allowing for varying levels of theoretical engagement. 
As a result, IPA and TA are methodologically coherent and complement each other by allowing individual insights to converge into shared themes while ensuring that broader themes remain grounded in phenomenological detail \cite{BraunClarke2012, Larkin2006}. 
We conducted IPA and TA concurrently and independently. 
The insights generated by each approach were subsequently discussed and integrated to develop a more comprehensive interpretative synthesis.

\subsection{Study Design and Workshop Protocol}\label{subsec:study-design}

We seek to understand how researchers engage with AI systems by examining (1) what prior experiences and expectations they bring from their everyday encounters with AI, (2) how they interact with AI systems during research tasks, (3) how they symbolically represent these interactions, and (4) how they articulate and reflect on these experiences in group discussions. To collect data on these different dimensions,
we designed a workshop structure that combined individual and group activities within a structured two-hour format. 
The planning and implementation of our workshops strictly adhered to established standards of research ethics (see \cref{apx:research-ethics} for more details).

Our interdisciplinary research team comprises backgrounds in human-computer interaction, psychology, design, cognitive neuroscience, computer science, machine learning, law, and research ethics. All authors are active users of such AI systems and share a professional curiosity about how researchers integrate AI into their work. These combined perspectives shaped our analysis toward technical, experiential, cognitive, and ethical considerations in interpreting participants’ accounts. Participants were recruited through institutional mailing lists, and only few were previously known to the research team.

We recruited between seven and eight participants per workshop, 
which was sufficient to generate diverse perspectives for group discussions while still ensuring that each participant received adequate attention from our two-person facilitation team. 
The group size also matched the capacity of our facility and allowed for individual activities alongside group work. Participants were recruited through multiple channels,
including social media platforms,
bulletin boards,
and institutional mailing lists,
using a poster that advertised our workshop series as `Human-AI Relational Dynamics in Qualitative Research,' 
and no compensation was provided.
Our goal was to ensure that participants were intrinsically motivated by the relevance of the topic to their professional lives---%
in line with the first part of our analytical framework,
which focused on understanding experiences that participants themselves considered significant in their lives.

We conducted three workshops between March and May 2025.
A total of 22 qualitative researchers with diverse disciplinary backgrounds,
age groups,
genders, 
and nationalities participated in our study.
These researchers' experiences with AI systems varied considerably,
ranging from first-time users to those employing sophisticated prompt-engineering techniques.
This heterogeneity was intended because it allowed us to observe a broad spectrum of human-AI interactions while acknowledging participants' prior experiences.

\begin{figure*}[t]
  \centering
  \includegraphics[width=\linewidth]{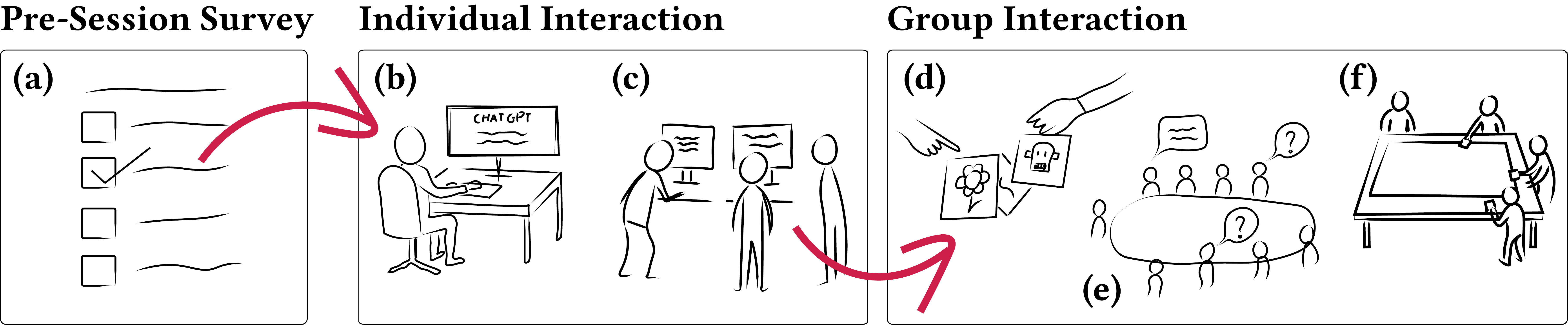}
  \caption{\textbf{Workshop structure.} 
  Progressing from individual to collective activities, 
  our workshops included (a) a pre-session survey capturing participants' initial perceptions of AI systems and their use in qualitative research; (b) individual hands-on interaction sessions with AI systems for qualitative-data-analysis tasks; (c) a peer observation activity where participants reviewed and reflected on one another's AI interactions; (d) an image selection and description exercise to elicit metaphorical representations of AI; (e) semi-structured group discussion exploring collective experiences and sense-making; and (f) a collaborative worksheet activity documenting shared insights. 
  }
    \Description{%
    Six-panel diagram illustrating our workshop methodology flow. Panel (a) shows a pre-session survey form with checkboxes. Panel (b) depicts a participant at a computer interacting with their assigned AI system. Panel (c) illustrates peer observation activity, where participants view one another's screens. Panel (d) illustrates an image selection exercise in which participants choose from visual metaphors. Panel (e) displays a semi-structured group discussion with participants seated in a circle. Panel (f) shows a collaborative worksheet activity with participants at a table using post-it notes. Red arrows connect the sequential phases from individual to collective activities.}
\label{fig:tasks}
\end{figure*}

As illustrated in \Cref{fig:tasks}, our workshops were structured into three sequential phases designed to explore participants' perceptions of and interactions with AI systems.

\begin{enumerate}[nosep,label= \emph{(\arabic*)},leftmargin=0em,itemindent=!]
    \item \emph{Pre-Session Survey.}\quad
    Before the workshop,
    participants completed a survey that included questions on their demographics as well as their attitudes, expectations, and concerns regarding AI systems (see \Cref{apx:survey-questions} for a complete list of the survey items along with their scales). We designed the survey drawing on the technology-acceptance framework \cite{davis1989perceived, venkatesh2000theoretical} to evaluate participants' overall intention to use AI. 
    Completing the survey also provided participants with an initial context for their engagement with the technology.
    \item \emph{Interactive Engagement with AI Systems (45m).}\quad
    In this phase,
    we assigned each participant to one of four AI systems (ChatGPT, Claude, Gemini, or Perplexity). We employed multiple AI systems to obtain richer perspectives, reduce the risk of model-specific biases, and enhance the reliability of our findings. 
    All participants were provided with fictional interview transcripts (designed to be contextually relevant and engaging) and tasked with generating a three-to-five-sentence summary and two research questions through interactions with their assigned system (see \cref{apx:fictional-interviews} for the verbatim task and transcripts).
    The last ten minutes of this phase were dedicated to a peer-review activity in which participants were encouraged to move around the room and 
    view the chatlogs on others' screens.
    This activity shifted focus from individual qualitative tasks to relational dynamics between participants and AI systems, and it additionally served as the group's first point of social contact.
    
    \item \emph{Roundtable Discussions and Exercises (75m).}\quad 
    The roundtable segment began with brief introductions and ground rules for respectful discussion,
    emphasizing the importance of turn-taking and active listening. 
    We then initiated an image-selection exercise. Drawing on the Zaltman Metaphor Elicitation Technique (ZMET) framework \cite{zaltman_coulter_1995_metaphor},
    we presented 34 images independently curated by the facilitators (see \cref{apx:visual-metaphors} for miniatures of the images on offer).
    Participants were instructed to choose one or more image(s) representing their perception of AI systems,
    based on their workshop interaction and any prior experiences.
    Using images as symbolic instruments helped participants articulate aspects of their AI interactions that might have been difficult to express directly.
    Following image selection,
    we guided a semi-structured discussion, allowing the conversation to flow naturally among participants while occasionally intervening using pre-structured probing questions to deepen the discussion (see \cref{apx:discussion-structure} for details).
    In this part of the roundtable segment, participants shared experiences,
    expressed opinions,
    and engaged with others' perspectives.
     In the final stage, participants were invited to split into two groups and complete worksheets with three sections:
     \emph{Feelings},
     \emph{Surprises}, 
     and \emph{Issues}. 
     Groups discussed and populated these areas using post-it notes and drawings while facilitators observed.
     The session concluded with a collective activity where all participants contributed to a fourth section on a whiteboard labeled as \emph{Actions},
     collaboratively developing concrete ideas and action plans based on their discussions about the role of AI in research contexts.
\end{enumerate}

\subsection{Analytical Framework: Individual and Social Dimensions}\label{subsec:analytical-framework}
The analytical process consisted of two complementary analyses---each of one conducted by one independent analyst---to emphasize different aspects of how qualitative researchers interact with and understand AI systems. One analysis was conducted at the \emph{individual level}, while the other was conducted at the \emph{social level}. Findings were then cross-discussed to create a unified narrative. 
On the \emph{individual level}, the analyst focused specifically on each participant's interactions with AI, their symbolic representations,
and personal reflections across different data segments (i.e., pre-session surveys, chatlogs, image selection and descriptions).
On the \emph{social level}, the researcher scrutinized how participants explained their experiences to others,
what concerns,
feelings,
and surprises emerged,
and which aspects of human-AI interactions were emphasized in a group setting (group-level data). 

Joining the individual and social dimensions of our analysis added nuance and depth,
revealing important relational dynamics unfolding across the different stages of our workshops.
For instance,
we could identify discrepancies between how participants engaged with AI (revealed in chatlogs) versus what they said they did (expressed in discussions),
or unearth tensions that only surfaced when participants discussed and compared experiences with one another.

\textbf{Individual Relational Dynamics.}\quad
We employed a multilayered,
sequential analysis designed to build a rich,
contextualized understanding of each participant's relational dynamics with AI. 
Drawing from the person-centered,
interpretative approach of \emph{Interpretative Phenomenological Analysis} (IPA) \cite{Smith2009, Smith2021},
we proceeded through distinct stages,
with each stage informing and adding complexity to the next.
IPA is a qualitative research approach that examines how people make sense of their lived experiences.
We adapted the IPA framework to examine not a single life event but an unfolding technological integration that profoundly impacts the participants' understanding of themselves,
their work,
and their place within rapidly changing academic and societal contexts.

To illuminate individual human-AI relational dynamics,
we documented the analysis of each participant using a custom digital archiving tool.
As the analyst examined different data sources from the workshops for each individual,
this tool helped create detailed case files that threaded together each participant's trajectory across various data streams.
We intentionally designed the analytical sequence to move from the most direct artifact (chatlogs) to progressively deeper layers of subjective context (pre-session surveys, image selection and descriptions).
This sequence enabled a progressive understanding of each individual case.
To ensure methodological integrity,
the analyst maintained separate reflexive memos throughout the process,
documenting emerging interpretations and preconceptions as they developed.

Each case was analyzed following a three-stage interpretive process. First, we conducted an inductive close reading of the chatlogs, identifying key moments, tensions, and shifts in the human-AI dialog. Second, we incorporated pre-session survey responses to situate these observations within each participant's attitudes, expectations, and concerns about AI. Third, we examined participants' image selections and descriptions to reveal the implicit frameworks and affective qualities shaping their AI interactions, grounding all interpretations in each participant's own articulated meanings. 

\textbf{Collective Sense-Making and Interaction Patterns.}\quad
In parallel to the IPA, 
we applied \emph{thematic analysis} \cite{Braun2022} to identify patterns within group-level data (group discussions). 
In doing so,
the participants’ perspectives,
as expressed during in worksheets and collective activities,
were also taken into account.
The analyst (1) transcribed the data from our recordings, (2) read and re-read the data to become familiar with the content, (3) took initial notes on potential themes that emerged during this process, (4) generated initial codes using both inductive and deductive approaches, and (5) discussed the initial codes with the other authors to refine them and resolved any disagreements. The analyst then (6) applied all agreed-upon codes to segments of group discussions, (7) grouped together codes with similar semantic meaning to form initial themes, refining and organizing them around semantic axes, and (8) constructed the narrative explaining the themes and their relationships.


\section{Findings}\label{sec:findings}

\begin{figure*}
    \centering
    \includegraphics[width=\linewidth]{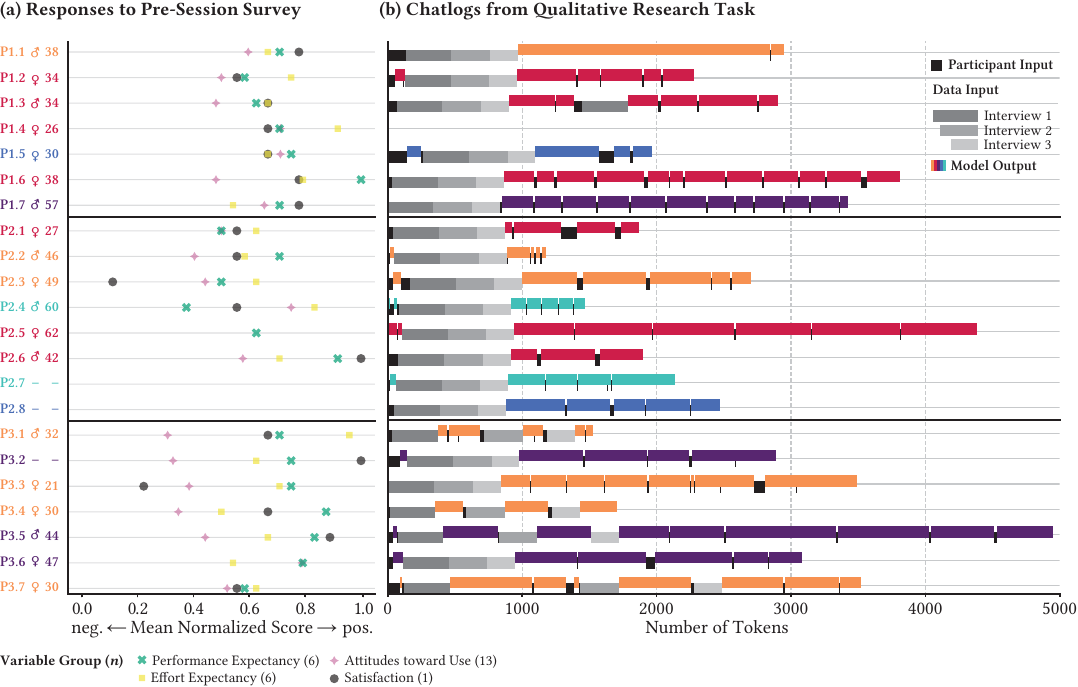}
    \caption{\textbf{Survey responses and chatlog data from the knowledge workers who participated in our study.} 
    We show participants' survey responses on core variable groups along with basic metadata~(a) as well as the turn-taking patterns in participants' chatlogs~(b), 
    Participant identifiers and chatlogs are colored by their dominant relational configuration (see \Cref{fig:teaser}). 
    See \cref{apx:data} for details.\newline 
    \emph{Data notes.} Two participants did not fill out the pre-session survey (\RTQBCI and \WYCXDJ), three did so only partially (\WGAJTB,\JPBRXV, \UWADUD), and one did not specify their gender or age (\LVAZBN).
    For one participant (\OQUJBJ), we were unable to retrieve the chatlog.
}
    \label{fig:chatlogs-survey}
    \Description{Two-panel figure displaying a subset of the data sources underlying our study: (a) Pre-session survey responses showing scores across 4 variables for 22 participants, with responses ranging from negative (0) to positive (1) on performance expectancy, effort expectancy, attitudes toward use, and satisfaction. (b) Horizontal bar chart showing chatlog token counts (0-5000) for each participant's interaction with AI systems. Participant IDs and chatlogs are colored by their dominant relational configuration.}
\end{figure*}

In this section,
we report how ontological and relational instability manifested across the individual and collective levels of our analysis. 
Participants are referenced using the letter `P', followed by the session number, a period, and the participant number (e.g., `P2.3' refers to the third participant in session two). 
\Cref{fig:chatlogs-survey} provides a visual summary of our survey and chatlog data. 
Through the examples below,
we elucidate how fundamental uncertainties about AI can make it challenging for individuals to hold a consistent relational stance in their interaction with an AI system.
We demonstrate how participants' attempts to frame AI using familiar categories---%
such as tool,
assistant,
and collaborator---%
repeatedly break down in real engagement.
Our research reveals a phenomenon that existing frameworks do not fully capture. 
Therefore,
we propose \emph{relational dissonance} as a new analytical construct to capture the dynamic implicit and relational dimensions of interactions between humans and anthropomorphic conversational agents. 

\subsection{AI Systems Between Tool and Other}\label{subsec:tool-other}

In our analyses, we found that our participants’ engagement with AI systems frequently exceeded purely instrumental interactions,
despite explicit utilitarian framing.
This insight emerged across our two complementary analytical approaches, when connecting all our data sources and examining them in different sequences. 

The worksheets filled out by our participants collectively highlighted that AI systems evoked profound and ambivalent responses---%
ranging from excitement to frustration, from
curiosity to suspicion, 
and from empowerment to loss of~agency. The range of responses we observe reflects not only a collection of mixed feelings but a fundamental difficulty in maintaining a stable relational stance toward inherently ambiguous systems.
Notably, in the worksheets,
some participants documented their surprise at witnessing peers ``humanize the agents'' or develop ``friend-like relationships'' with AI.
This surprise also reveals an implicit assumption that others maintain a purely instrumental orientation toward AI,
demonstrating how the assumed norms of AI interactions (`using' AI as a `tool') diverge from the actual relational configurations that unfold through conversational exchanges with AI.

\GMNEXL,
for instance,
framed AI as a powerful tool:
“I would draw a calculator if I knew how to draw… it is similar to using a calculator, a powerful calculator.”
\GMNEXL's survey showed positive views on AI's research utility but strong discomfort with emotional engagement and overdependence concerns.
This tension suggests the instrumental approach attempts to maintain boundaries against a technology they worry might become more than a tool.
In the participant’s chatlog,
at first glance,
their engagement with Perplexity (a proprietary model) appears instrumental,
involving directive prompts.
However,
this effort to control the interaction goes beyond using a tool,
as the participant maintained boundaries that would not be needed with a more predictable tool (``...first I want you to make a full report. This report should contain ${\text{... '')}}$.
Here,
dissonance exists between the articulated `tool' frame and the reality of engaging with the AI system.

A similarly instrumental framing manifested differently for \MJCEWA, who engaged with Claude (Opus 3.5):
``AI is the car that I am driving.
I only understand the wheel right now...
need to be aware of the snake on the road.''
The participant then remarked,
``If I don't master this tool within the next five years,
I could be obsolete as a professional.''
\MJCEWA's survey responses showed positive assessments of AI utility coexisting with concerns regarding domination,
overdependence,
and societal consequences.
Hence,
the tool-mastery stance reflects both personal and general concerns about maintaining control amid technological change.
However,
\MJCEWA's chatlog revealed distinctly relational exchanges,
such as offering praise like ``Excellent work!''
and managing the AI’s limitations by requesting it to ``indicate that you don't know.''

An entirely different relational configuration appeared with \WGAJTB, who positioned AI interaction as a technical competency to master:
``I think learning prompt engineering is like learning any other skill,
like riding a bike.
It's difficult and frustrating in the beginning,
and you fall,
and you know there are false starts and but then when you get it,
you can be a lot faster.''
When analyzing fictional qualitative data with the AI system (Perplexity),
the participant asked,
``What if I took a constructivist approach?''
This question subtly repositioned AI from an instrument requiring skillful operation to an intellectual collaborator capable of engaging with theoretical perspectives.
The AI adjusted to this shift by responding
``Here's how you could frame constructivist research questions...,''
which transformed the interaction from prompt engineering to a theoretical discussion.
When the AI's outputs fell short of expectations,
the participant provided feedback:
``The previous summary was better,''
and suggestions:
``I would focus on these.''
The interaction became an active iterative calibration,
similar to a pedagogical exchange.
\WGAJTB's survey answers indicated enthusiasm about AI's research utility,
but also strong discomfort about emotional engagement with AI, being judged by it,
and losing agency.
The chatlog reflected this tension, as the initial frame of technical mastery momentarily gave way to collaborative inquiry, then shifted into a pedagogical hierarchy.

Viewing the chatlog data alongside participants' verbal accounts revealed that the initial prompts are typically aligned with participants' prevailing conceptualizations of AI.
We observed explicit attempts to establish an interactional frame from the outset,
likely as prompting strategies to control both the output and relational dynamics.
Some participants (\GMNEXL, \BLQUQS, \LVAZBN, \YSDDVC) employ role-assigning prompts such as ``You are an experienced qualitative data researcher'' (\GMNEXL)
or ``You are an excellent qualitative researcher;
you have the ability to read the transcript and provide insightful findings from the transcribed interviews'' (\LVAZBN).
Others attribute roles implicitly (e.g., \MJCEWA, \GYGLNT, \RTQBCI, \UWADUD), using phrases like
``You are conducting research on supervisory relationships in PhD work'' (\MJCEWA).
These opening prompts represent attempts to establish an ultimately untenable relational configuration:
assigning AI to a specific expert role while simultaneously maintaining directorial control over the expertise.
Our analysis indicates that these initial ontological and relational frames rarely hold during the interaction. 
This may be attributed to the essential affordances of anthropomorphic conversational agents:
Even when explicitly constrained, these systems continually generate adaptive responses that engage with and actively reconfigure their users’ initial relational expectations.
In this context,
control becomes less about directing a tool and more about navigating an evolving relational dynamic that neither party fully determines. 

The examples given above demonstrate how relational dissonance, i.e., the discrepancy between how participants conceptualize AI systems and what transpires as they interact with them, manifests in our study setting. To explore this phenomenon in greater depth,
we proceeded to identify both the predominant relational configurations and the critical moments that shifted relational dynamics during participants' interactions with AI.

\subsection{Oscillating Relational Configurations}\label{sec:results-oscillation}

To systematically analyze the relational dynamics in our data,
we first inductively identified five typical relational configurations (see \Cref{fig:teaser}a).
We derived these configurations by mapping the dominant relational dynamics surfacing in each session and distilling them into concise summaries.
Additionally, we selected specific adjectives and verbs that best captured the relational qualities of each unique exchange (e.g., characterizing one participant as a `Guiding mentor training a dependent novice' and another as a `Curious researcher co-creating with a supportive consultant').
By synthesizing these diverse engagements found in our data, we defined the following recurring categories:
\begin{enumerate}[nosep]
    \item \emph{Director}: directive, task-oriented prompting (e.g., P3.4: ``summarize this article to 3-5 sentences, particular focus on sentence with // mark'');
    \item \emph{Trainer}: assumes pedagogical authority to train or steer the system (e.g., P1.6, issuing evaluative feedback to shape the output: “The previous summary was better. Rewrite.”);
    \item \emph{Partner}: engages in collaborative iteration where ideas are co-constructed (e.g., P3.5, establishing a social opening “Hello Gemini!” before framing the workflow as a joint effort: “I will give you the texts and I need you to give me suggestions for the summaries.” );
    \item \emph{Student}: seeks guidance, exhibiting receptivity or accepting reframing (e.g., P1.5, explicitly asking their own analysis to be verified “analyse if my answer is same as you would get”, prompting the system to adopt a pedagogical stance: “Your summaries are mostly accurate, but I would refine them...”);~and
    \item \emph{Consumer}: accepts AI responses passively with minimal involvement (e.g., P2.7, delegating the task via a minimal “please do a summary” without specifying constraints or focus, and accepting the response as the findings).
\end{enumerate}
These categories primarily capture participants' relational orientations, rather than the roles of AI systems or participants’ conceptual understanding of them.
While they serve as a necessary scaffold for our analysis,
interpreting them as static labels does not suffice to capture the relational dynamics we observed.
Hence, we do not regard these categories as fixed relational positions;
rather,
they provide a high-level framework for understanding the dominant interaction styles of the participants in our data.

The most notable finding from this mapping is the dominance of control-oriented positions,
with seven participants categorized as directors (\GMNEXL, \DDMHLY, \EZFYQG, \DPYKRI, \SFWPFB, \SRXTXD, \YSDDVC) and seven as trainers (\GYGLNT, \MJCEWA, \OQUJBJ, \WGAJTB,  \BLQUQS, \JPBRXV, \KOLLDL).
This suggests that our participants primarily engaged with AI systems by establishing hierarchical,
control-oriented relationships---%
in line with their professional norms.
However,
during image selection and description tasks,
even participants with controlling approaches characterized systems through language suggesting ontological uncertainty,
using phrases such as ``something ominous'' (\EZFYQG),
``dark people'' (\RZWGHO),
and ``emotionless kind of thing'' (\KOLLDL).
These findings indicate that control-oriented approaches reflect participants' attempts to stabilize the ontological status of AI via familiar hierarchical structures in interactions. 
Strikingly, control-oriented approaches shifted very quickly when AI systems or the participants themselves introduced adaptive or social overtures, e.g., 
by offering explicit positive feedback.
Such feedback exchanges occurred among trainers (\GYGLNT, \MJCEWA, \BLQUQS),
partners (\ZIKDKJ, \LVAZBN, \TMZZKH),
and students (\RZWGHO).
Trainers that began with clear hierarchical stances would suddenly offer appreciation:
``Excellent work!'' (\MJCEWA),
``Thank you, this is helpful.'' (\GYGLNT),
``I like your words...'' (\BLQUQS)
Consumers (\GUKOJD, \RTQBCI) and directors did not demonstrate explicit reciprocity,
but this absence reflected their relational configurations, rather than a lack of dynamics.
In particular, consumers tended to approach interactions with curiosity and disengaged upon obtaining information,
and directors continuously worked to maintain hierarchical boundaries,
sometimes by fully forgoing conversational engagement.

The dynamic nature of these interactions prompted us to seek a more granular analytical tool to understand how the roles of humans and AI systems shift during human-AI interactions, 
and we initially sought to adopt Lindgren's framework \cite{Lindgren2025} describing human-AI functional and social roles.
However, although this framework effectively categorizes dominant relational orientations,
it was not designed to capture role shifts within a given interaction.
Hence, we adapted the framework by adding new relational roles to their existing categories and mapping the distinct relational turning points.
Through this technique, we were able to reveal continuous role oscillations that occurred between conversational turns in ongoing human-AI interactions,
and what initially appeared to be stable categories dissolved into continuous relational negotiations:
A single participant might shift from directive control to collaborative partnership to student-like deference within minutes,
often in response to relational shifts initiated by the AI system.
Thus, individual relational configurations should be viewed as momentary constellations, rather than as fixed positions.

For example,
\LVAZBN started their interaction with ChatGPT 4o by explicitly assigning it the role of a `qualitative researcher,'
thus positioning AI as a specialist.
Initially,
the AI behaved proactively and collaboratively.
However,
when the participant challenged the system's output,
the AI's relational stance became more adaptive and deferential.
Rather than simply correcting its response,
the system validated the challenge:
``You're absolutely right to challenge that.''
Shortly after this,
the participant corrected the AI again and again received a similar validation, accompanied by appreciation:
``Excellent observation!''
Hence,
what began as a structured approach to prompting was transformed into a dynamic exchange in which the participant actively refined the analysis provided by their AI system through a negotiated,
iterative dialogue.
When the participant provided positive feedback by saying
``I like this idea,''
the system responded,
``I am glad to hear that!'',
and the interaction progressed into an exchange of mutual appreciation. 

The process of negotiation and mutual adaptation,
illustrated in the example,
repositions the system's output not as a static product of technology but as a direct outcome of the relational configurations between humans and AI.
Our attempts to impose discrete phases on fundamentally continuous relational adjustments reveal the limitations of segmented approaches.
The most important finding of this analysis is not which categories individuals or AI systems belong to,
but rather that they do not remain in any category long enough for such classifications to holistically capture the human-AI relational space.
This highlights the need to develop approaches that are better suited to capture the relational dynamics that emerge when humans interact with anthropomorphic conversational agents. 
\emph{Relational dissonance} takes one step toward this goal.  
We now proceed to show how it manifests in a social~setting. 

\begin{figure*}
    \centering
    \includegraphics[trim={0.5cm 1.5cm 1cm 0.5cm}, width=\linewidth]{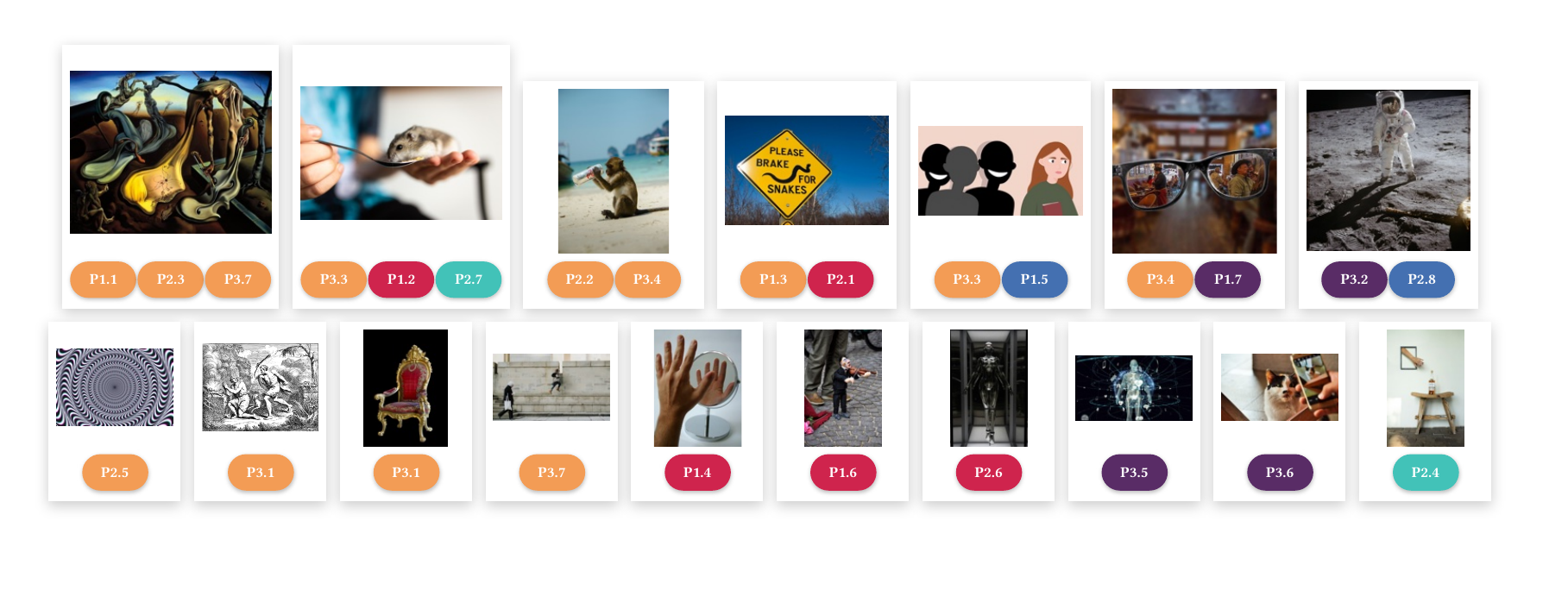}
    \caption{\textbf{Visual metaphors selected by the knowledge workers who participated in our study.} 
    Images are scaled by the number of participants who selected them, 
    and participants who selected an image are represented by badges and colored by their dominant relational configuration (see \Cref{fig:teaser}). 
    Four participants selected two visual metaphors instead of just one metaphor (\DPYKRI, \SFWPFB, \SRXTXD, \YSDDVC). 
    For the full set of images presented to participants during the image-selection exercise, see \Cref{fig:metaphors} in the Appendix.}
    \label{fig:metaphors-selected}
    \Description{Gallery presenting the images selected by the participants in our study. 
    The figure shows two rows of images that are scaled by the number of participants who selected them. 
    Each image is annotated with the identifiers of the participants who selected it, who are represented by badges in the color of their dominant relational configuration. 
    }
\end{figure*}

\subsection{Collective Sense-Making of Relational Dissonance}\label{subsec:collective-sense-making}

Relational dissonance is not limited to participants’ individual interactions with AI systems but also extends into the collective sphere. Within the group interactions, 
participants actively engaged in a sense-making process, 
drawing on shared social representations and categories of meaning to articulate and negotiate their experiences. 
Through these discursive sense-making processes, 
the group collectively recognized, 
contested, and reframed the tension between cognitive, 
conscious reflections and relational, 
often less conscious dynamics. 
In this context, 
relational dissonance becomes not only an individual psychological experience but also a phenomenon shaped by collective sense-making. 
Epistemic assumptions, 
professional norms, and the relational dynamics specific to academic communities play an important role in how relational dissonance is negotiated in collective discourse.

This discursive sense-making process becomes particularly visible in the way participants used metaphorical language to describe AI---%
encapsulated by the visual metaphors they chose, reproduced in \Cref{fig:metaphors-selected}. 
In particular, participants often employed language of depth to construct their relationship with AI (\emph{Theme 1: Depth of AI}). 
Metaphors of depth, 
speaking of “going deeper” (\JPBRXV), 
“under the hood” (\MJCEWA), 
or having "the potential to sort of dig deeper or unmask or something within us” (\TMZZKH), position AI as layered, 
immersive, and multidimensional. 
Such metaphors are associated with creativity and intellectual expansion, as one participant noted: “I agree that it does aggregate knowledge, but still, it might kind of... And it's a creative leap” (\YSDDVC). Yet, 
dissonance is evident in counter-positions that frame AI as superficial, rather than profound: “they are like parrots, 
they are not doing something deeply” (\WYCXDJ). 
These metaphors resonate with Lakoff and Johnson’s \cite{lakoff2008metaphors} account of conceptual metaphors, 
where knowledge is spatialized along a deep/shallow axis, 
and with Sweetser’s insight that abstract domains such as thought or emotion are mapped through spatial schemas \cite{bodwell1993eve}. 
In this context, 
“going deeper” (\JPBRXV) functions as a cognitive tool, shaping how participants framed AI’s epistemic potential. 
By framing AI as deep or shallow, 
participants were also participating in a broader cultural negotiation over the epistemic agency of machines. 
As highlighted in recent work on the cultural transformation mediated by AI, 
human sense-making increasingly occurs within an ecology where machines not only mediate but also generate cultural forms \cite{brinkmann2023machine}. Therefore, 
depth metaphors can be read as markers of how participants make sense of the transmission of knowledge across human–machine boundaries. 
For academics, 
these metaphors take on particular salience, 
as the pursuit of depth is closely tied to their practices of knowledge production.

Closely connected to metaphors of depth is the theme of uncertainty and the unknown (\emph{Theme 2: Uncertainty and Unknown}). 
On the one hand, participants expressed opacity, 
doubt, 
or confusion about how AI works and what it might produce. 
These expressions framed AI as unpredictable and difficult to grasp: “they are these dark people” (\RZWGHO),
“it’s kind of foreign” (\RZWGHO), 
or simply “I’m not even understanding” (\MJCEWA). Participants also highlighted unpredictability in outcomes: “so I somehow understand the basics of cats. So how they behave and how can I touch or not touch and how do I approach it. But then I think that they are wild animals inside. So they are unpredictable.” (\UWADUD) or adopting a trial-and-error stance: “let’s see where it gets me” (\ZIKDKJ). 
On the other hand, 
participants articulated moments of transparency, suggesting that AI could mirror one’s own input or beliefs: “a mirror is somewhat what you see,
and it’s like what you feed. 
So whatever you think about yourself or what you believe in yourself, you see.” (\OQUJBJ).
This counter-position illustrates dissonance in the tension between opacity and transparency in participants’ collective sense-making.
These constructions of uncertainty resonate with broader discourses on algorithmic agency and inscrutability. 
For example, Pasquale described AI systems as part of a “black box society,” 
where operations remain hidden and uncontrollable \cite{pasquale2015black}. 
Burrell identified different forms of opacity in machine learning, from technical complexity to social inaccessibility \cite{burrell2016machine}, 
while Eiband et al.\ highlighted how users perceive AI as driven by “hidden processes” \cite{eiband2018bringing}. While academic professional practice is oriented toward making knowledge visible and transparent, here, participants found themselves grappling with systems that resist such clarification. 
This tension underscores how relational dissonance is shaped by the pervasive experience of opacity in human–machine relations.

Importantly, 
the collective sense-making highlighted in \emph{Theme 1}  and \emph{Theme 2}  is not only epistemic but also relational, shaping how participants positioned themselves in relation to AI. 
As part of this positioning, participants showed a tendency to `other' AI, 
treating it as alien, non-human, or purely machinic (\emph{Theme 3: Othering}). 
Participants described AI as “an emotionless kind of thing” (\KOLLDL), or “just a glorified chatbot” (\MJCEWA). 
At the same time, 
the discussions revealed counter-positions, 
where participants anthropomorphized AI, attributing to it human-like traits, 
intentions, or even identities \cite{de2017people}. 
For example, some participants highlighted the communicative ability of AI, 
suggesting that “they are people because I think they can communicate with me like another human would.” (\RZWGHO).
These contrasting framings illustrate how relational dissonance was enacted through oscillations between distancing and humanizing the machine, 
reflecting uncertainty about where to position AI within relational categories. 
Accordingly, 
participants assigned a variety of relational roles to AI (see \cref{subsec:tool-other}), 
which were negotiated collectively and ultimately shaped the collective sphere of relational dissonance. 
This collective dimension is particularly relevant in the academic context, 
where relational positioning directly affects how knowledge is produced and validated.


\section{Discussion}\label{sec:discussion}

Our workshops with qualitative researchers revealed a fundamental tension in human-AI interactions:
a discrepancy between people's explicit framing of AI systems and the relational dynamics that manifest during their actual engagement with these systems.
We term this divergence \emph{relational dissonance,}
shaping a novel analytical construct to understand the implicit dynamics that ultimately determine the processes and outcomes of knowledge work.
In this section,
we first clarify the role of relational dissonance in human-AI interactions and its implications for the instrumental use of AI in knowledge work.
We then explore the concept of \emph{relational transparency} to address possible methodological, epistemic, and personal risks resulting from human-AI relational dynamics.

\textbf{Relational Dissonance as a Central Element of Human-AI Interactions.}\quad
Prominent HCI theories \cite{Nass1994, Nass2000}
propose that people develop and apply technology-specific relational scripts when interacting with computers.
However,
our data consistently show that anthropomorphic conversational agents introduce relational complexities that transcend script-based conceptualizations.
Instead of simply executing relational scripts,
our participants interacted with their AI systems in dynamic and relational negotiations.
The relational dissonance resulting from these negotiations is shaped by the interplay of two primary factors: 
(1) the inherent \emph{ontological and relational ambiguity} of AI systems themselves, and (2) multiple \emph{individual and contextual factors},
including the participants' professional identities, their individual expectations, and 
societal narratives and expectations about AI.

Relational configurations that arise during human-AI interactions are based on a fundamental \emph{asymmetry of consequences}. This asymmetry is mirrored in the chatlog data (\Cref{fig:chatlogs-survey}): Based on participants' brief inputs, AI generates lengthy outputs, which highlights the scale of the AI's contribution to the task at hand. Yet, despite this potentially major influence, 
the emerging relational dynamics carry no tangible stakes for the AI system,
whereas the outcomes have real-world implications for the human.
The concerns about agency loss,
overdependence,
or erosion of cognitive skills,
explicitly reported by participants (\cref{subsec:tool-other}),
indicate an intuitive recognition of their precarious position in human-AI interactions.
In the current implementations of AI systems, 
the entire responsibility for validation and accountability falls on the individual,
who must navigate relational configurations that emerge without their full control.
This \emph{unilateral responsibility} in an asymmetric relational dynamic contributes significantly to the persistent tension that constitutes the essence of relational dissonance by making relational instability consequential.
When knowledge workers bear the sole responsibility for their outcomes,
each relational shift becomes something they might need to justify or defend. The interaction requires their relational engagement to produce knowledge, but it simultaneously demands vigilance toward the same system they must rely on.
As our data shows,
people often find themselves trying to extract stable results---by defining roles,
correcting outputs,
demanding transparency---%
from a process that is structurally unable to provide that stability.

Although generative unpredictability (the capacity to engage in human-like dialogue, 
surprise, 
and provoke)
is precisely what sets these systems apart from traditional computers,
the same features also create patterns of engagement that go beyond participants’ articulated relational scope, as the continuous role oscillations in \cref{sec:results-oscillation} demonstrated.
The constant overflow of affective and relational engagement in human-AI interactions makes relational dissonance inevitable and central to how these interactions unfold.

Although there is increasing recognition of the need for new relational frameworks \cite{Brailsford2025},
much of the existing HCI research on AI remains concentrated on either the user's cognitive state \cite{Xu2024, Vereschak2024, Raees2025} or the system's performance or features \cite{Tsiakas2024, He2025}. 
Our findings suggest that the most critical interactions occur at the interface between the user and the AI system.
Thus,
we suggest that the primary focus when analyzing human-AI interactions should be on the relational dynamics developing at that interface.
This shift in analytical focus is critical for studying relational dissonance,
which does not reside solely in either the user or the system but rather emerges from the implicit tensions that develop between the two parties during their interaction.

\textbf{The Entanglement of Relational and Analytical Work.}\quad
Contemporary discourse on AI in knowledge work is dominated by `tool use' narratives.
Industry leaders urge professionals to `master this new tool' \cite{webb_altman_ai_tools_2025},
universities offer prompt-engineering courses or guides on `how to use AI' \cite{webb_nyu_ai_tools_2025}. 
These framings position AI as a sophisticated but controllable instrument that can be integrated into existing workflows with proper training and methods.
The emphasis on prompting strategies,
best practices, 
and optimization techniques delivers the message that successful human-AI interaction is primarily a matter of technical proficiency.

Our data demonstrates that a purely instrumental use of anthropomorphic conversational agents without relational engagement is untenable within the context of qualitative research and knowledge work more broadly. As shown in \cref{subsec:tool-other}, instrumental boundaries inevitably give way to relational engagement.
The more participants sought to manage the relational dissonance and its underlying tensions,
the more \emph{relational work} they had to perform.
Participants’ attempts to impose relational structures,
define roles,
manage emotional responses,
or correct outputs can all be viewed as instances of such relational work.

The dynamics observed within our small groups of qualitative researchers encapsulate a larger phenomenon unfolding at a societal scale.
Our data shows how professionals skilled in methodological reflexivity engage in significant relational work during their interactions with AI systems.
For millions of people who interact daily with anthropomorphic conversational agents, relational dynamics may become even more complex.
Each interaction becomes an unscripted relational exchange that individuals must navigate,
often without the conscious awareness or conceptual frameworks necessary to articulate their own experiences. 

The impact of relational dynamics on knowledge work is profound.
As participants in our workshops uploaded data directly to AI systems,
the chat interface became their primary analytical space.
This merging of analytical space and conversational medium means that knowledge work, 
such as qualitative research, 
becomes inseparable from the relational dynamics of the conversation itself.
Consequently,
the quality and nature of knowledge produced through human-AI interactions now also depend on how researchers navigate human-AI relational dynamics. 
This often remains unacknowledged in academic and professional discourse.

Qualitative research practice has long been characterized by scholars' continuous examination of their own interpretive influence on their data. 
When using AI systems in their work, 
qualitative researchers must maintain this reflexivity not only toward themselves but also toward AI systems as contributors to the interpretive process. 
However, the relational complexities within the human-AI interface make the continuous reflexivity required by existing qualitative frameworks harder to enact,
with implications for the development of future qualitative methods.

\textbf{A Shift Toward Relational Transparency.}\quad
The relational dynamics occurring at a conversational interface are not directly controlled by any single actor,
whether human or artificial.
Recognizing this limited control is necessary for understanding how relational configurations diverge from users' initial relational frameworks, 
and it enables novel approaches to the design of interfaces, systems, and policies.
The persistence of relational dissonance in our data suggests that addressing it requires systemic intervention---not to eliminate the built-in tension, but to surface the burdensome, unacknowledged relational work it generates in contexts with real professional stakes. 
Hence, we propose a shift in focus toward \emph{relational~transparency}: While technical explanations might partially illuminate how training on massive datasets and fine-tuning processes contribute to creating ontologically ambiguous systems,
it is insufficient for understanding the relational dynamics at the human-AI interface. Rather than attempting to control or eliminate relational engagement---%
an approach that is particularly ineffective with anthropomorphic conversational agents---,
relational transparency seeks to make relational dynamics and dissonance visible and comprehensible to users, designers,
and regulators. 
The aim is to reveal how relational configurations evolve, transform,
and impact the outcomes of knowledge work---%
and perhaps also human-AI interaction more broadly. 

Implementing relational transparency involves embedding accountability within the system design,
offering context-aware visibility of relational states that adapts to varying situations,
and providing support to help users recognize relational dissonance.
This approach distributes responsibility across the socio-technical infrastructure,
rather than placing it solely on individual users,
and it provides designers and policymakers with a foundation for concrete action.

Putting relational transparency into practice requires a multilayered approach that addresses (1) the human-AI interface, (2) system design, and (3) broader policy implications. First, at the level of the \emph{human-AI interface},
systems can expose relational dynamics and reveal relational dissonance through real-time feedback or post-interaction digests.
This may include alerts or visualizations when the relational configurations shift.
Such features will assist users in navigating implicit dynamics reflexively. 
Second, at the level of \emph{system design},
relational transparency can involve logging relational metadata,
tracking relational configurations,
and detecting linguistic patterns. This way, the system will assume responsibility for capturing and disclosing relational shifts,
rather than expecting users to independently monitor relational dynamics.
And third, at the level of \emph{AI governance}, 
relational transparency may be a desirable regulatory goal that is insufficiently recognized by existing regulatory frameworks.  
While emerging AI regulation increasingly mandates logging and system-level accountability measures for anthropomorphic conversational agents, 
the regulatory instruments in place overwhelmingly analogize these agents to tool-like machines, ignoring their nature as relational artifacts. 
Explicitly including relational dynamics in system audits may be a first step toward incentivizing relational transparency by design, and it could eventually lead to \emph{relational nutrition labels} summarizing AI systems' relational inclinations. 
However, more interdisciplinary research is needed on how AI governance can effectively integrate human-AI relational dynamics. 

\section{Limitations}\label{sec:limitations}

\textbf{Conceptual Limitations.}\quad 
While this study conceptually distinguishes relational dissonance from related constructs such as \emph{cognitive dissonance} and the \emph{intention-behavior gap,}
our empirical evidence is mostly qualitative,
based on discrepancies between explicit accounts and implicit dynamics identified across our multiple data sources. Future research could focus on developing more direct measures to quantitatively substantiate relational dissonance. 
For example,
specific conversational moves---%
such as praise, greetings, evaluative feedback, or pushback---%
could form the basis of a coding scheme for creating observable markers of shifts in relational stance, 
and divergences between people's explicit articulations or initial exchanges with an AI system and what their interaction traces subsequently reveal could be operationalized as indicators of the resulting structural tension.
We~further acknowledge that,
as researchers already invested in a relational perspective,
our analytical lens inevitably shaped our interpretation.
Although our findings are based on specific participant accounts and chatlogs,
it is important to view our framework as just one interpretive lens among many.
Moreover,
as relational dissonance partially operates outside conscious awareness,
our methodology captures its observable manifestations, rather than the complete psychological process of relational dissonance.
The value of this construct lies in illuminating the continuous micro-negotiations between explicit framings and implicit dynamics,
which cannot be fully captured by HCI taxonomies that assume stable~roles.

\textbf{Methodological Limitations.}\quad 
Our workshop methodology introduced constraints that may have shaped the phenomenon itself.
The artificial setting,
prescribed research task,
and peer observation might have actively generated the tensions we documented, rather than simply revealing preexisting dynamics.
Moreover,
the 45-minute interaction period might have been insufficient 
to distinguish between the initial friction of learning a new interface and the emergence of persistent relational dissonance, 
especially for participants bringing little or no experience with a particular system.
Additionally,
the workshop's explicit focus on \emph{Human–AI Relational Dynamics} and our task instruction to work ``in partnership with your AI'' may have primed participants to attend to relational aspects they might have otherwise navigated unconsciously.
This raises a critical question for future research:
Does relational dissonance emerge primarily under conditions of scrutiny, or does it operate continuously in everyday use? 
The ontological and relational ambiguity of anthropomorphic conversational agents \emph{theoretically} suggests that relational dissonance should be a ubiquitous phenomenon, 
but our approach cannot exclude that it \emph{practically} arises only under specific conditions of scrutiny. 
Furthermore, our study only documents the emergence of relational dynamics but not their long-term development---%
and it is likely that more complex forms of relational dissonance emerge through sustained natural use with real stakes and self-directed objectives. 
As our methodology captures only brief interactions during a specific period, 
elucidating the temporal dimensions of relational dissonance will require additional longitudinal and ethnographic research. 

\textbf{Empirical Limitations.}\quad 
Our findings emerge from a specific population of qualitative researchers operating within the professional norms of knowledge production,
agency,
and methodological reflexivity.
We cannot claim that the relational dynamics we document are universal or representative of the general population.
The tensions between tool-use framing and relational engagement might be especially pronounced in our study population,
which has professional stakes in interpreting qualitative data.
As entirely different modes of relational engagement may arise during everyday interactions with AI systems, 
generalizing our findings to a general audience is not straightforward. 
Therefore,
our empirical contribution should be understood as documenting how relational dissonance manifests in knowledge-work contexts where professional integrity and epistemic authority are at stake,
rather than identifying universal dynamics of human-AI interaction. 
However, our proposal that relational dissonance may be useful for understanding human-AI interaction more broadly extends beyond what the empirical evidence in this paper demonstrates.
Whether our construct applies similarly across different populations, settings, and tasks remains a question for future research.

\section{Conclusion}\label{sec:conclusion}

Our study addresses a critical gap in understanding human-AI interactions in the context of knowledge work.
Through our workshops,
we showed how implicit relational dynamics with AI systems shape how knowledge is produced,
validated,
and understood.
The quality and the production of intellectual output increasingly depend on navigating relational configurations that operate below conscious awareness, 
and we presented relational dissonance as a novel analytical construct for characterizing these implicit human-AI relational dynamics.
Our multilayered workshop protocol provides a replicable method for studying human-AI interactions from a relational perspective, 
and our qualitative study offers nuanced insights into human-AI relational dynamics in the qualitative-research context.

Beyond knowledge work,
anthropomorphic conversational agents are now being rapidly deployed at a large scale.
These systems typically exhibit certain relational characteristics,
often referred to as `personalities,'
which are either embedded within them or customizable by users.
As discussed in this paper,
at the conversational interfaces between humans and AI systems, 
relational dynamics emerge, often with unpredictable outcomes.
The deployment of such systems in broader populations effectively recruits millions of users as participants in an undeclared, uncontrolled experiment,
with potentially serious consequences.
Companies that provide access to anthropomorphic conversational agents intervene in human relational processes,
largely without sufficient understanding of the dynamics that occur when people engage with these systems.
The reactive cycle of product updates and feature recalls reflects the significant cost of treating relational complexity as a secondary issue, rather than as a primary consideration.

The concept of relational dissonance provides essential analytical vocabulary to capture what happens when humans interact with AI systems.
Future work can build on this construct, for example, by constructing and evaluating research prototypes that embody relational transparency while simultaneously developing a more comprehensive theoretical framework to guide this new design paradigm.
Finally, we encourage the development and implementation of relational-transparency measures as a foundation for the more responsible design of AI systems.
As anthropomorphic conversational agents are becoming ubiquitous,
addressing relational dissonance is not merely an academic concern but an essential component of human-centered technological development.

\begin{acks}
This work was supported by the Jenny and Antti Wihuri Foundation. We gratefully acknowledge the Aalto Center for Qualitative Research (Qual+) and the House of AI at Aalto University for providing the platform that enabled this research. CC further acknowledges the research environment provided by ELLIS Institute Finland.
\end{acks}

\balance
\bibliographystyle{ACM-Reference-Format}
\bibliography{chi2026_references}
\balance

\clearpage
\appendix

\section{Data Availability Statement}

Five types of data were collected in this study: (1) pre-session surveys, (2) individual chatlogs, (3) images selected as visual metaphors, (4) recordings of group discussions, and (5) group worksheets. Pre-session surveys and chatlogs (see \Cref{fig:chatlogs-survey}) were reviewed to ensure that no direct or indirect identifiers were present; these anonymized datasets are shared as part of the supplementary material of this paper. 
In the survey data, nationalities were censored to protect participant privacy. 
For an overview of the selected images, see \Cref{fig:metaphors-selected}. 
Group discussions and group worksheets contained direct identifiers and other personal data of the participants and are therefore not included in the supplementary material. 
However, they can be accessed under controlled conditions for the sole purpose of validating the findings. Derived data, including researcher-created annotations, is also available in the same controlled-access~repository.
\section{Research Ethics}\label{apx:research-ethics}

This study was reviewed and approved by the research ethics committee from the authors' organization. No significant risks were identified during the review process. As personal data from participants was collected, the study adhered to all applicable data-protection regulations. Each participant received a privacy notice describing the processing of their personal data, along with a consent form. Participation was voluntary, and participants were reminded that they could withdraw from the study at any time without consequence. No compensation was provided. To minimize privacy concerns related to the use of generative AI tools, participants used anonymous accounts supplied by the researchers. They were explicitly instructed not to enter any personal data during their interactions with the chatbots. Participants were also reminded that during the workshop, other participants would review their conversations with the AI tools. 

\section{Study Details}\label{apx:data}

In this appendix, we provide key details related to the setup of our study, sharing

    (1) the mapping between the human-readable participant identifiers used in the paper and the random identifiers used in the study itself (\cref{apx:participant-identifiers}),
    (2) the survey questions and item scales used in the pre-session survey (\cref{apx:survey-questions}),
    (3) the precise wording of the fictional qualitative-research task to be addressed by participants through human-AI interactions (\cref{apx:fictional-interviews}),
    (4) the precise versions of the AI systems used by our participants (\cref{apx:ai-systems}),
    (5) the images presented during the metaphor-elicitation exercise (\cref{apx:visual-metaphors}), and
    (6) the structure underlying the group discussions (\cref{apx:discussion-structure}).


\subsection{Participant Identifiers}\label{apx:participant-identifiers}

The mapping of participant numbers to the identifiers used in the source data is shown in \Cref{tab:participants}.

\begin{table*}[h]
    \centering
    \caption{Mapping of participant numbers to the participant identifiers used in the source data. Participant numbers follow the pattern (Session).(Participant), where participant numbers are assigned in alphabetical order of the participant identifiers in each session.
    }\label{tab:participants}
    \begin{tabular}{rlrlrl}
    \toprule
        \bfseries Number &  \bfseries Identifier&\bfseries Number&\bfseries Identifier&\bfseries Number&\bfseries Identifier\\\midrule
         \GMNEXL& GMNEXL&\BLQUQS&BLQUQS&\DPYKRI&DPYKRI\\ 
         \GYGLNT&GYGLNT&\DDMHLY&DDMHLY&\LVAZBN&LVAZBN\\
         \MJCEWA&MJCEWA&\EZFYQG&EZFYQG&\SFWPFB&SFWPFB\\
         \OQUJBJ&OQUJBJ&\GUKOJD&GUKOJD&\SRXTXD&SRXTXD\\
         \RZWGHO&RZWGHO&\JPBRXV&JPBRXV&\TMZZKH&TMZZKH\\
         \WGAJTB&WGAJTB&\KOLLDL&KOLLDL&\UWADUD&UWADUD\\
         \ZIKDKJ&ZIKDKJ&\RTQBCI&RTQBCI&\YSDDVC&YSDDVC\\
         &&\WYCXDJ&WYCXDJ\\
         \bottomrule
    \end{tabular}    
\end{table*}
\subsection{Survey Questions and Item Scales}\label{apx:survey-questions}

In the following, we reproduce the questions participants answered in our pre-session survey, organized by their variable group in \Cref{fig:chatlogs-survey}, along with the scales on which they could provide their answer (see \Cref{subsec:study-design}).  
In addition to these questions, we also asked participants to self-report their age, nationality, and gender as basic demographic information. 
All questions were optional. 
An anonymized CSV file of all answers (with nationalities censored to protect privacy) is provided in the supplementary~material.

\paragraph{Warm-Up Questions}
\begin{enumerate}
    \item Do you conduct qualitative research? [Yes $\mid$ No]
    \item Have you ever used, or currently using, an AI system to conduct your qualitative research? [Yes $\mid$ No] (Adapted from \cite{apolinario2019determinant})
    \item Which of the following AI systems have you frequently used (at least once a week) to conduct your qualitative research? [Selection + free text]
\end{enumerate}

\paragraph{Performance Expectancy [Five-Point Likert Scale from Strongly Disagree to Strongly Agree] (Adapted from \cite{davis1989perceived})}

\begin{enumerate}
    \item Using AI systems in my qualitative research would enable me to accomplish tasks more quickly.
    \item Using AI systems would improve my qualitative research performance.
    \item Using AI systems in my qualitative research would increase my productivity.
    \item Using AI systems would enhance my effectiveness in qualitative research.
    \item Using AI systems would make it easier to do my qualitative research.
    \item I would find AI systems useful in my qualitative research.
\end{enumerate}
\paragraph{Effort Expectancy [Five-Point Likert Scale from Strongly Disagree to Strongly Agree] (Adapted from \cite{davis1989perceived})} 
\begin{enumerate}
    \item Learning to operate AI systems would be easy for me.
    \item I would find it easy to get AI systems to do what I want them to do.
    \item My interaction with AI systems would be clear and understandable.
    \item I would find AI systems to be flexible to interact with.
    \item It would be easy for me to become skillful at using AI systems.
    \item I would find AI systems easy to use.
\end{enumerate}
\paragraph{Social Influence [Five-Point Likert Scale from Strongly Disagree to Strongly Agree] (Adapted from \cite{apolinario2019determinant})}
\begin{enumerate}
    \item People close to me would approve of AI systems.
    \item My friends would approve of AI systems.
    \item My boss would approve the use of AI systems.
\end{enumerate}
\paragraph{Facilitating Conditions [Five-Point Likert Scale from Strongly Disagree to Strongly Agree] (Adapted from \cite{apolinario2019determinant})}
\begin{enumerate}
    \item I have the necessary technical preconditions for using AI systems.
    \item I possess the technical know-how to utilize AI systems.
\end{enumerate}
\paragraph{Attitudes toward Use [Five-Point Likert Scale from Strongly Disagree to Strongly Agree] (Adapted from \cite{nomura2006measurement})} 
\begin{enumerate}
    \item I would feel uneasy if AI systems really had emotions.
    \item Something bad might happen if AI systems developed into living beings.
    \item I would feel relaxed talking with AI systems.
    \item I would feel uneasy if I was given a job where i had to use AI systems.
    \item If AI systems had emotions I would be able to make friends with them.
    \item I feel comforted interacting with AI systems that have emotions.
    \item The word ``AI system'' means nothing to me.
    \item I would feel nervous operating an AI system in front of other people.
    \item I would hate the idea that AI systems were making judgments about things.
    \item I feel that if I depend on AI systems too much, something bad might happen.
    \item I would feel paranoid talking with an AI system.
    \item I am concerned that AI systems would be a bad influence on children.
    \item I feel that in the future society will be dominated by AI systems.
\end{enumerate}
\paragraph{Behavioral Intention [Five-Point Likert Scale from Strongly Disagree to Strongly Agree] (Adapted from \cite{curcuruto2009fiducia})} 
\begin{enumerate}
    \item If I have the opportunity, I intend to use the AI systems to carry out my qualitative research.
    \item I foresee that, if it is available to me, I will use the AI systems in my qualitative research.
    \item Being able to access the AI systems at work, I intend to use them as much as possible.
\end{enumerate}
\paragraph{Voluntariness of Use [Seven-Point Likert Scale from Strongly Disagree to Strongly Agree] (Adapted from \cite{moore1996integrating})} 
\begin{enumerate}
    \item My superiors expect me to use AI in qualitative research.
    \item My use of AI in qualitative research is voluntary (as opposed to required by my superiors or job description).
    \item My boss does not require me to use AI in qualitative research.
    \item Although it might be helpful, using AI in qualitative research is certainly not compulsory in my job.
\end{enumerate}
\paragraph{Engagement [Seven-Point Likert Scale from Never to Always] (Adapted from \cite{schaufeli2018ultra})}
\begin{enumerate}
    \item At conducting qualitative research with AI, I feel bursting with energy. 
    \item I am enthusiastic about my qualitative research with AI.\\
    \emph{Note: Due to a formatting error, this item was presented on a five-point Likert scale (also from Never to Always). Although this inconsistency is a limitation, prior cross-cultural validation studies have shown that the scale functions robustly with both five-point and seven-point formats \cite{schaufeli2018ultra}.}
    \item I am immersed in my qualitative research with AI. 
\end{enumerate}
\paragraph{Satisfaction [Ten-Point Likert Scale from ``Not at all satisfied'' to ``Completely satisfied''] (Adapted from \cite{dolbier2005reliability})}
\begin{enumerate}
    \item Thinking specifically about your qualitative research with AI, indicate how satisfied you feel:
\end{enumerate}

\subsection{Fictional Qualitative Research Task}\label{apx:fictional-interviews}

\paragraph{Task Description.}~\phantom{\hspace{1em}}

The following excerpts are from interviews conducted with three doctoral students at the same research institution. 
Each was asked the same question:
\emph{"How is your relationship with your supervisor?"} 

Goals of the Exercise

In partnership with your AI, 

a)	Review the interviews and craft a 3-5 sentence summary that highlights the most intriguing elements that stood out to you.

b)	Formulate 1-2 research question(s)- Optional.

\paragraph{Interview Transcripts}~\phantom{\hspace{1em}}

\subparagraph{Participant 1}~\phantom{\hspace{1em}}

Um, my supervisor, right? (He looks nervous, keeps avoiding my eyes.) Umm, this isn't going to be like, published anywhere... I mean, with my name and stuff, right? (I reassure him: "No, everything will be anonymized and aggregated into the results.") Good, okay… (He pauses for a long moment.) She’s polite… friendly.  But then… sometimes, she barely seems interested in my work? Like, she quickly scans through my drafts and goes, “Hmm, interesting,” then asks me to—uh—do other stuff, like update databases or proofread her articles. I remember, she even asked me to organize data for a project completely outside my research. She said I had a "good eye" for sorting things out. (He laughs sarcastically.)

I keep expecting her to, like, actually read my stuff, you know? Or at least say something that makes sense about it, but… I don't know. She nods like she's paying attention, but then says something completely random... And when I try to get clarity, she's like, “You'll figure it out,” as if it's obvious or something. It’s just—honestly... (He pauses.)

I end up working late on stuff that's not even mine, while my own research just…  Sometimes I wonder—maybe I'm doing this wrong, like I should know… or just handle it myself? (He sighs quietly, looking down.) I don't know.

And everyone keeps saying things like, “You are lucky, she's so pleasant to work with,” and it makes me feel kind of bad for even mentioning it. Because yeah, she's nice but… (He rubs his forehead) it doesn't actually help. And now I'm in my third year... Everyone's talking about how many articles they have, who's getting published, who's behind. And I've got, like, one paper under review, but I need more, you know? I'm honestly panicking a bit. (He pauses, visibly anxious.) Sorry, I think I rambled too much… I hope this sort of answered your question?
(He seems uncertain, looks around nervously throughout. Maybe feeling uncomfortable for voicing these concerns?)\\

\subparagraph{Participant 2}~\phantom{\hspace{1em}}

Honestly? My relationship with my supervisor is a bit complicated. (She crosses her arms.) He's energetic, always buzzing around, networking, you know the type. At first, I was kind of relieved—someone like that... maybe actually helpful. But soon I noticed my ideas, my frameworks, my visuals popping up in his drafts. Under his name... (Shakes her head.)
First time this happened, I was sort of polite. I just said, "Hey, shouldn't I be a co-author on this?" He gave me this look, you know, like I'm being unreasonable, and then said something like, "Why are you turning this into something emotional?" That word (I think she’s referring to the word “emotional”) just hit me wrong. 

So, um, I went to admin, filed a formal complaint, explained everything clearly. They said they'd handle it. But nothing happened. Not even a follow-up email.  At that point, I realized, alright, this is how it is. But I'm documenting everything now, because I'm not letting this slide again.

You know, uh, I'm in my late 30s, I've been through stuff. I'm pretty good at setting boundaries, usually, but this… Especially because every time I try to draw a line, he smiles politely, maybe says something slightly inappropriate—never openly rude, just... unsettling—and somehow everyone else laughs it off. And I'm left feeling... You know, it's like you are talking and no one's hearing.

Honestly, it's beyond just authorship now. I'm not some endless resource… I'm preparing to escalate again, even if it means a huge fight with administration or changing supervisors entirely. (Exhales deeply.) Sorry, this turned into a rant. (Laughs)
(Throughout the interview, she frequently pauses as if carefully choosing her words.)\\


\subparagraph{Participant 3}~\phantom{\hspace{1em}}

Hmm... We don't talk much. So basically, I send him my draft and then two weeks, three weeks, no reply. Suddenly, an email comes, very short, just saying, "Looks okay, keep going." I guess this is feedback? (He laughs.) Not really sure.

Before, I used to think, ah, maybe my questions too simple, or maybe too stupid… like I’m troubling him, you know? So basically, I stopped asking already. Now, I just… check everything myself, double-check, triple-check, before I send. (He adjusts his glasses) Don’t want him to think I cannot handle the work.
But then, maybe he wants me to be independent? (He laughs lightly.) Maybe that's good? Or... I, maybe I'm... I try not to take it personally, but it's hard. If I make a mistake, I fix it myself first because I'm worried if I mess something up, he'll notice me for the wrong reason. 

Sometimes I see other students with their supervisors, and I feel a bit jealous because they get real advice. But maybe less pressure for me, less micromanaging? I don't know. (Another laugh here) So yeah, that's how it is. 
(I noticed it's almost like he laughs whenever he says something significant.)

\subsection{AI Systems Used by Participants}\label{apx:ai-systems}

\paragraph{Workshop 1 (24 March 2025)}\phantom{\hspace{1em}}

\begin{itemize}[--]
\item \GMNEXL: Perplexity Sonar \emph{(Perplexity’s in-house model.)}

\item \GYGLNT: ChatGPT-4o

\item \MJCEWA: Claude 3.5 Haiku

\item \RZWGHO: ChatGPT-4o

\item \WGAJTB: Perplexity Sonar 

\item \ZIKDKJ: Claude 3.5 Haiku

\item \OQUJBJ: Gemini 2.0 Flash
\end{itemize}

\paragraph{Workshop 2 (29 April 2025)}\phantom{\hspace{1em}}

\begin{itemize}[--]
    \item \BLQUQS: ChatGPT-4o

\item \DDMHLY: (First interaction) Perplexity Sonar, (Second interaction) ChatGPT-4o

\item \EZFYQG: Gemini 2.0 Flash

\item \GUKOJD: Gemini 2.0 Flash

\item \JPBRXV: ChatGPT-4o

\item \KOLLDL: Perplexity Sonar

\item \RTQBCI: ChatGPT-4o

\item \WYCXDJ: ChatGPT-4o
\end{itemize}

\paragraph{Workshop 3 (23 May 2025)}\phantom{\hspace{1em}}

\begin{itemize}[--]
\item \DPYKRI: Gemini 2.5 Flash

\item \LVAZBN: ChatGPT-4o

\item \SFWPFB: Perplexity Sonar

\item \SRXTXD: ChatGPT-4o

\item \TMZZKH: Gemini 2.5 Flash

\item \UWADUD: ChatGPT-4o

\item \YSDDVC: Gemini 2.5 Flash
\end{itemize}

\paragraph{Notes on the Choice of AI Systems.}

We relied on AI systems provided by third parties to ensure the ecological validity of study and accommodate resource limitations.
Relying on third-party AI systems required us to adapt to evolving free-tier limits and pricing structures. Following the initial session, usage caps on Claude's free tier, along with similar restrictions with Perplexity, significantly disrupted interactions. 
To maintain the integrity of our dataset, we limited the use of Claude and Perplexity to a smaller group of participants. These practical constraints resulted in a higher proportion of ChatGPT and Gemini users in the final dataset. 
\subsection{Images Provided for Selection as Visual Metaphors}\label{apx:visual-metaphors}

\begin{figure*}[!h]
    \centering
\includegraphics[width=0.9\linewidth]{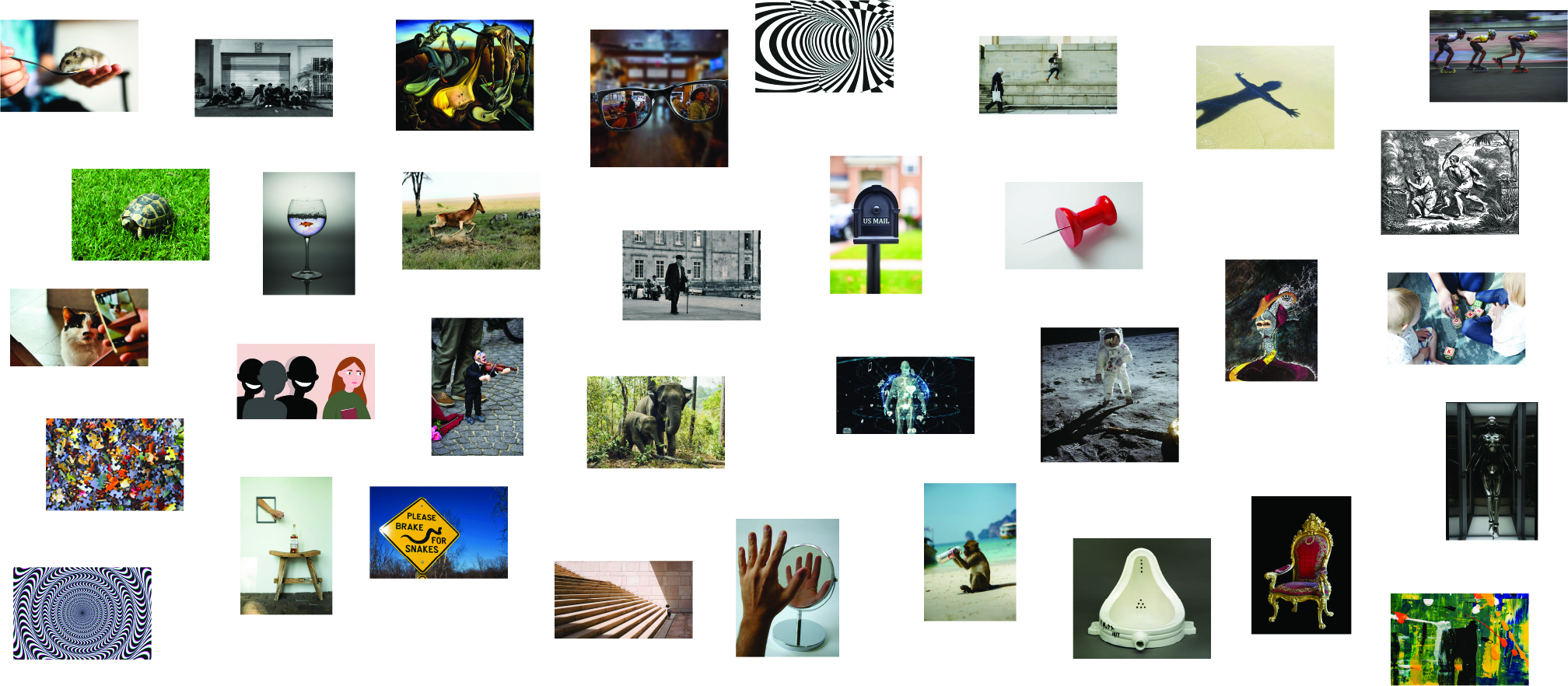}
    \caption{\textbf{Visual metaphors used in the image-selection exercise.} 
    }
    \label{fig:metaphors}
    \Description{%
    Grid of 34 diverse images used as visual metaphors during the image-selection exercise. Images include: abstract patterns, animals, human representations, technology, various objects, nature scenes, other symbolic imagery, such as a road sign "Please Break for Snakes", a hand reflecting from a mirror, and a throne. Images are arranged in no particular order.
    }
\end{figure*}

\Cref{fig:metaphors} reproduces miniatures of the 34 visual metaphors presented to participants in the image-selection exercise. 
The images were sized to fit standard printer paper and arranged on a table in no particular order.

\subsection{Group Discussion Structure with Facilitating Questions}\label{apx:discussion-structure}

We used the following structure for the third component of our workshop (roundtable discussions and exercises).
 
\begin{itemize}
    \item \textbf{Rules (5 min)}
    \begin{itemize}
        \item Communicate the rules of the workshop:
        \begin{itemize}
            \item Speak clearly, one at a time, to facilitate interaction.
            \item Keep contributions brief.
            \item Collaborate positively and constructively: listen to and respect others' opinions.
            \item Feel free to express your own opinion and experience, adding to what has already been said.
            \item Feel empowered to change your mind without fear of seeming inconsistent.
            \item Use imagination, associations of ideas, and analogies.
        \end{itemize}
        \item Inform participants that the round table discussion will be recorded.
        \item Emphasize: each contribution is crucial due to the uniqueness of participants’ experiences. There may be different perspectives, but there are no right or wrong answers; all have the "right to exist" in the group. Participants are encouraged to discuss among each other, focusing on the interaction with AI.
    \end{itemize}

    \item \textbf{Presentations \& Post-it (5 min)}
    \begin{itemize}
        \item Spell out random code, name, and brief description of role/department.
        \item Write name on Post-it.
        \item Choose one or more images that represent your idea of AI, having in mind what you have done so far (do not overthink). Materials: printed images.
    \end{itemize}

    \item \textbf{Image Discussion (15 min)}
    \begin{itemize}
        \item Why did you choose that/those picture(s) to represent your idea of AI? If you are in the picture(s), where are you, who are you? (10 min)
        \item Any questions, surprises, or curiosities regarding each other’s choices? (5 min)
        \item Pictures are then placed aside. 
    \end{itemize}

    \item \textbf{Worksheets \& Group Reflection (25 min)}
    \begin{itemize}
        \item Participants in subgroups complete worksheets naming key surprises, feelings, and concerns in using AI for qualitative research (3 boxes).
        \item \textbf{Surprises}: Unexpected or noteworthy moments during AI interactions.
            \item \textbf{Feelings}: Emotional reactions and personal responses triggered by AI.
            \item \textbf{Issues}: Challenges, difficulties, concerns, or unresolved aspects in the session or when using AI in qualitative research.
            \item Guiding questions: What is the problem? Who is involved? Who is affected? Why is this happening? Why do people care?
        \item Participants collaborate and discuss answers, using sketches if helpful.
    \end{itemize}
    \item \textbf{Action Building (10 min, with the board)}
    \begin{itemize}
        \item \textbf{Actions} – Practical steps, potential applications, or future ideas sparked by AI in qualitative research.
        \item Guiding questions: What do you want to achieve? How can this be solved? How will it be done?
    \end{itemize}

    \item \textbf{Closing (10 min)}
    \begin{itemize}
        \item Is there anything else important that we have not discussed? Did anything not come up?
    \end{itemize}

    \item \textbf{Post-session Evaluation \& Chocolate (5 min)}
    \begin{itemize}
        \item Short evaluation questions.
        \item Chocolate.
    \end{itemize}
\end{itemize}

\section{Themes Resulting from Thematic Analysis}\label{apx:thematic-analysis}

The thematic analysis sketched in \cref{subsec:analytical-framework} resulted in the themes detailed in \cref{tab:thematic-analysis}. These themes form the basis of the discussion presented in \cref{subsec:collective-sense-making}.

\begin{table*}[ht]
    \caption{\textbf{Codes used in thematic analysis.}}
        \label{tab:thematic-analysis}
    \centering
    \begin{tabular}{p{0.2\linewidth}p{0.235\linewidth}p{0.235\linewidth}p{0.235\linewidth}}
    \toprule
         \bfseries Theme &\bfseries Definition &\bfseries Positive examples & \bfseries Negative example\\\midrule
         Theme 1: Depth of AI & 
         Use of metaphorical language that presents AI as deep, layered, immersive, multi-dimensional or having spatiality. The opposites emphasize AI is shallow or one-dimensional.&	
         ``going deeper''
(P2.5); ``under the hood'' (P1.3); ``the potential to sort of dig deeper or unmask or something within us'' (P3.5).& ``they are like parrots, they are not doing something deeply'' (P2.8).\\\midrule
    Theme 2: Uncertainty and Unknown&Expressions of opacity, doubt, or confusion about how AI works or what it might do, unpredictability. The opposite are expressions of transparency.&``they are these dark people'' (P1.5); ``it’s kind of foreign'' (P1.5); ``I’m not even understanding'' (P1.3).&``a mirror is somewhat what you see and it’s like what you feed. So whatever
you think about yourself or what you believe in yourself, you see.'' (P1.4).\\\midrule
    Theme 3: Othering&Treating AI as non-human or ``other,'' personified or described as alien or agentic. The opposite is considering AI as human or having human attributions.&``an emotionless kind of thing'' (P2.6);  ``just a glorified chatbot'' (P1.3).&``they are people because I think they can communicate with me like another human would.'' (P1.5).\\
    \bottomrule
    \end{tabular}
\end{table*}

\end{document}